\documentclass[10pt]{article}
\usepackage{amsfonts}
\usepackage{amssymb,amsmath}

\newcommand{\beqn}{\begin{equation}}
\newcommand{\eeqn}{\end{equation}}

\def\br{\begin{eqnarray}}
\def\er{\end{eqnarray}}
\def\brn{\begin{eqnarray*}}
\def\ern{\end{eqnarray*}}
\def\er{\end{eqnarray}}
\def\beq{\begin{equation}}
\def\eeq{\end{equation}}

\def\vp{\varphi}

\def\a{\alpha}
\def\b{\beta}

\def\e{\boldsymbol {e}}

\def\bv{\mathbf{v}}
\def\bbr{\mathbf{r}}
\def\bpsi{\boldsymbol{\psi}}
\def\bR{\boldsymbol{R}}
\def\dbpsi{\boldsymbol{\dot{\psi}}}
\def\ddbpsi{\boldsymbol{\ddot{\psi}}}
\def\dddbpsi{\boldsymbol{\dot{\ddot{\psi}}}}

\begin{document}
\title{Equations of motion for a (non-linear) scalar
field model as derived   from the field equations}

\author{\thanks { email kaniel@math.huji.ac.il;\quad  itin@math.huji.ac.il}
 Shmuel Kaniel$^{(a)}$ and  Yakov Itin$^{(a,b)}$\\
 {\tt $^{(a)}$Institute of Mathematics, Hebrew University of
  Jerusalem }\\
  {\tt $^{(b)}$Jerusalem College of Technology, Jerusalem, Israel}}


\newcommand{\bi}[1]{\bibitem{#1}}
\date{\today}

\maketitle
\begin{abstract}
 The problem of derivation of the equations of motion
 from the field equations is considered.
 Einstein's field equations
have a specific analytical form: They are linear in the
second order derivatives and quadratic in the first order
derivatives of the field variables. We utilize this
particular form and propose a novel algorithm for the
derivation of the equations of motion from the field
equations.  It is based on the condition of the balance
between the singular terms  of the field equation.    We
apply the algorithm to a nonlinear Lorentz invariant
scalar field model. We show that it results in the Newton
law of attraction between the singularities of the field
moved on approximately geodesic curves. The algorithm is
applicable to the $N$-body problem  of the Lorentz
invariant field equations.
\end{abstract}
 PACS: {04.25.-g, 45.50.Dd, 95.10.Ce}
\maketitle
\section{Introduction}          

General Relativity (GR) is unique among the class of field
theories in the treatment of the equations of motion. When the
particles are modeled by the singularities of the field, their
motion is completely determined by the field equation.
 By comparison, in classical electrodynamics, the equations of motion of
matter sources are postulated independently from the field
equations. A precise analysis of the basis assumptions of
classical electrodynamics was given recently in the premetric
axiomatic framework of Hehl and Obukhov \cite {Birkbook},
\cite{Itin:2004qr}. This way, two independent assumptions have to
be postulated separately from the Maxwell field equations: (i)
{\it The agent of interaction}
--- the Lorentz force expression,  and (ii) {\it The agent of
inertia} --- the mass-times-acceleration term.

In GR, the field equations are nonlinear so a superposition of
solutions does not satisfy the field equations. Consequently, the
motion of sources cannot be completely  independent of the field
equations. Moreover, it  was shown already in the early days of
GR that the motion of a point-like particles is  determined by the
field equations and should not be postulated separately. This
result was achieved in two  essentially different ways:

(i) Due to Einstein and his collaborators  \cite{EIH0} --- 
\cite{infeld2} (see also \cite{Weyl} --- \cite{havas}),  
 the equations of motion of the point-like
particles were derived from the vacuum  field equations
\begin{equation}\label{int1}
R_{ij}-\frac 12 g_{ij}R=0\,.
\end{equation}
Point-like particles are represented  by the singularities  of the
gravitational field, i.e, of the solutions of (\ref{int1}).

(ii) Fock  \cite{Fo}, see also \cite{Petrova},   considered
particles as represented by a suitable energy-momentum tensor
$T_{ij}$. Correspondingly,  the equation of motion  was derived
from the full Einstein field equation
\begin{equation}\label{int1x}
R_{ij}-\frac 12 g_{ij}R=T_{ij}\,.
\end{equation}
Papapetrou \cite{Pap}, does not assume a specific form
 of the energy-momentum. Thus his approximation method
 is fairly general and, in principle, can be carried out
 to arbitrary high orders.

In both models,  a suitable approximation scheme was used so that
the equations of motion of massive particles, were approximated by
 geodesic curves.

The Einstein-Infeld-Hoffmann (EIH) equations were used for
high order PPN approximations of the $N$-body problem 
\cite{Car} --- 
\cite{Blanchet:1998vx}, for
 calculation of the gravitational radiation reaction force \cite{And},
 and
 for describing the motion of a gyroscope \cite{Schiff},
 \cite{Anderson:2005vj}. For the geodesic line approximation
 of the motion of bodies of a sufficiently small size and mass,
see \cite{E-G} and the reference given therein.

The EIH-approach was used to establish arguments for the plausibility
 of the geodesic postulate.
In particular,  Sternberg \cite{St} proved the geodesic postulate
in a general form for a system that incorporates also Yang-Mills
fields.

The aim of our paper is different. In this article, we
adopt Einstein's idea of describing  point-like particles
by the singularities of a field.  Then, some   natural
questions can be considered:
\begin{itemize}
\item[(i)] What analytical properties of the Einstein field
equations are used in order to derive the proper equations  of
motion for singularities?
\item[(ii)] What parts of the field equation have to be
identified with the  {\it agent of inertia} and the {\it  agent of
the force}?
\item[(iii)]
What other nonlinear field equations lend themselves to
derivability of the equations of motion?
\end{itemize}
 In this article we submit partial answers.
 First we define a suitable class of the field
equations. These are linear in the second order derivatives and
quadratic in the first order derivatives of the field  variables.
We also require the equations to be Lorentz invariant. We show
that the linear part serves as the agent of inertia while the
quadratic part is the agent of the  force.

We suggest a nonlinear scalar Lorentz invariant field model.
 There is an exact static solution to this equation, which is singular
 at $N$ points.
 This solution does not imply interaction between singularities,
 it is non physical. Indeed, for the Newtonian gravity as formulated
 by a field theory,
 $\triangle \,\phi=0$, a similar exact solution\footnote
 {We use a system of units with $c=1\,,G=1$.}
 $\phi=\sum_i  {m_i}/{r_i}$ exists.
 It is rejected by an additional axiom, which
 is Newton's law of attraction. Also for  Maxwell
 electrodynamics, the potentials satisfy (in the Lorentz gauge)
 the wave equation $\square\, A=0$. An exact static solution is
 $A=\sum_i  {q_i}/{r_i}$. Also this solution is non physical
 (it does not describe the field of free non-constrained
sources) because it contradicts  Coulomb law. In both
examples, one cannot distinguish between the quasi-static
and dynamic solutions by the field equations.

 We do not pre-postulate the law of interaction.
 It is derived from the field equation. For our model,
the non-physical exact static solution is rejected, the
same way the quasi-static solution was rejected in
Newton's and Maxwell's theories.  Instead, an approximate
solution is constructed. When our algorithm
is applied to this  solution,  it yields  the Newton-type law
 of attraction.

Our algorithm  can be briefly described as follows.
 \begin{itemize}
\item[(i)] Start with a nonlinear Lorentz invariant field equation.
Let its static spherical symmetric solution (exact or approximate)
be singular at a point. Due to  Einstein's description, this is
the field of one  particle.
\item[(ii)] The field of a particle moving with a constant velocity
is constructed by a  Lorentz transformation  of the static field.
\item[(iii)] The field of a particle moving  on a curved trajectory\footnote
 {The bold symbols denote 3-dimensional
vectors.} $\bpsi(t)$ is constructed by a Lorentz
transformation with an instantaneous velocity $\dbpsi(t)$.
\item[(iv)] The field of $N$ particles is taken to be, approximately,
a superposition of the fields of the single particles.
\item[(v)] When this field of $N$ particles is inserted
into the field equation,  two singular expressions emerge: one
from the linear part and the other from the quadratic part. By
equating the highest order terms of these expressions the
strength of the singularity of the solution is reduced. The
corresponding balance equation  yields Newton's law of attraction
between the singularities.
\end{itemize}

We apply this algorithm to a nonlinear Lorentz invariant scalar
field model. In this
context, the Newton law of attraction is derived.

 The paper is organized as followed:
 In section 2, we construct a nonlinear scalar Lorentz invariant
 field model. It shares some analytical properties of the Einstein
 field equation.
  In section 3, exact static solutions are presented.
The solution with $N$ static singularities does not allow to
describe the interaction. We treat it as non-physical.
 In section 4, we construct a non-static solution with $N$
 singularities.
 In section 5, we show that the balance between the
 singular contributions yields, in the first order, the Newton law
 of attraction.
 In section 6, we present a qualitative technical summary
  of our algorithm.
 In section 7, we give a brief outlook of the proposed alternative
 algorithm and its possible extensions.

\section{A nonlinear scalar model}
\subsection{Pointwise singularities in GR}
In vacuum, the Einstein field equation can be written as
\begin{equation}\label{Eins}
R_{\mu\nu}=0\,.
\end{equation}
Here and in the sequel, the Greek indices refer to the coordinates,
$\mu, \nu,\cdots=0,1,2,3$,  while the  Roman indices denote the
specific singular point of the solutions, $ i,j, \cdots=1,\cdots,N$.
 From an analytical point of view, (\ref{Eins}) is a system of ten
equations for ten components of the metric tensor
$g_{\mu\nu}$. All the equations are:
\begin{itemize}
 \item [(i)] Hyperbolic nonlinear PDE of the second order;
\item [(ii)] Linear in the second order derivatives
of the metric tensor;
\item [(iii)] Quadratic in its  first order derivatives.
\end{itemize}
The field of a pointwise source is described by the Schwarzschild
solution of (\ref{Eins}), which nonzero components are:
\begin{itemize}
\item [(i)] Smooth for $r\ne 0$;
\item [(ii)] Singular at the origin $r=0$;
\item [(iii)] Have the Taylor expansion of the form
\begin{equation}\label{Taylor}
g_{\mu\nu}\sim 1+\frac mr+C_1\left(\frac
mr\right)^2+C_2\left(\frac mr\right)^3+\cdots\,.
\end{equation}
\end{itemize}
The positive parameter $m$ in (\ref{Taylor}) is interpreted as a
mass of a pointwise source located at the origin of the
coordinates $\bbr=0$, while the constants $C_1,C_2,\cdots$ are
dimensionless.

 Due to the translation invariance of (\ref{Eins}),
the mass $m$ can be located at an arbitrary point,
 $\bbr=\bbr_0$. Moreover, it was proved, see \cite{F-B}, that the system
(\ref{Eins}) has a smooth solution with $N$ singularities located
at the points $\bbr=\bbr_i(t)$. Although,  even for $N=2$, an exact
 form of such a
solution  is not available, the leading terms of its Taylor
expansion can be calculated. The main output of the EIH analysis,
is the fact that the functions $\bbr_i(t)$ describe
 the proper
motion of small bodies. Particularly, in the first order
approximation,  the functions  $\bbr_i(t)$ satisfy the Newton
equation of motion with the Newton force of attraction at the
right hand side. The higher order approximations yield the
geodesic equation.
\subsection{Nonlinear scalar model}
 In order to clarify how the field equation itself can determine
 an interaction between singularities,
 we will study  a scalar field model which shares the special analytical
 properties of the Einstein equation mentioned above.

 Let a scalar field $\vp=\vp(x)$ be given on a flat Minkowskian
$4D$-space with a metric
$\eta_{\mu\nu}=\eta^{\mu\nu}=diag(1,-1,-1,-1)$.
 We consider a nonlinear scalar field  equation
\begin{equation} \label{s-eq}
\eta^{\mu\nu}\left(\vp_{,\mu\nu}- k\vp_{,\mu}\vp_{,\nu}\right)=0\,,
\end{equation}
which is  equivalent to
 \begin{equation} \label{s-eq1}
 \square\, \vp-k\left(\dot{\vp}^2-||\nabla \vp||^2\right)=0\,.
\end{equation}
Here $k$ is a dimensionless constant, it's value  will be
discussed in the sequel.

The commas denote partial derivatives $\vp_{,\mu}=\partial
\vp/\partial x^\mu$,
 the wave operator is $ \square=
 \partial^2/\partial t^2-\partial^2/\partial x^2-
 \partial^2/\partial y^2-\partial^2/\partial z^2$,
 $\nabla $ is the spatial gradient operator, the scalar product
 $<\cdots>$ and the corresponding norm $||\cdots||$ are Euclidean.

 In our approach, the scalar model (\ref{s-eq}) is used only as a
 simplified illustration  to the way  an interaction between
 singularities can emerge in  GR.
 The model, however, is interesting by
 itself.
 In particular, the numerical evolution of a scalar field model
 similar to the one under consideration
 was recently studied  in \cite{WM}.
\section{Exact static solution}
\subsection{Linear equation}
In order to derive a suitable exact solution to (\ref{s-eq1}), we start with
 a special case $k=0$. For this value,  the
equation (\ref{s-eq}) is  linear
\begin{equation}\label{w-eq}
\square \,\vp=\eta^{\mu\nu}\vp_{,\mu\nu}=0\,.
\end{equation}
Its unique static, spherically symmetric,
asymptotically zero solution is
\begin{equation}\label{w-sol1}
\vp=\frac mr\,.
\end{equation}
Note that this function  is  smooth
 for $r\ne0$ and  singular at the origin of the coordinate system. It is
natural to interpret such a  solution as a field of a pointwise
particle   with a mass (or a charge) equal to $m$ located at the
origin.
 The translational invariance of (\ref{w-eq}), yields that the
 singularity can be located at an arbitrary point $\bbr_0$
\begin{equation}\label{w-sol1x}
\vp=\frac {m}{R}\,,
\end{equation}
where
\begin{equation}\label{w-sol2x}
\bR=\bbr - \bbr_0\,,\qquad R=||\bR||\,.
\end{equation}
The linearity of the field equation (\ref{w-eq}) yields the
existence of a static  solution which is singular at $N$ distinct
points
\begin{equation}\label{w-sol2}
\vp=\sum^N_{i=1}\frac {m_i}{R_i}\,,
\end{equation}
where
\begin{equation}\label{w-sol2xx}
\bR_i=\bbr - \bbr_i\,,\qquad R_i=||\bR_i||\,.
\end{equation}
This solution approaches zero at infinity and is approximately
spherically symmetric in small neighborhoods of every
singularity. For long distance from the set of the singular
points,
 the full spherical symmetry is reinstated
 \begin{equation}\label{w-sol2xy}
\vp=\frac M r\,, \qquad M=\sum^N_{i=1}{m_i}\,.
\end{equation}
 We interpret (\ref{w-sol2}) as a field generated by
a static configuration of $N$ particles of masses $m_i$ which are
located at the points $\bbr_i$. Since the solution (\ref{w-sol2})
is static, an interaction between the particles is absent.
\subsection{A nonlinear equation}
 Let us derive now  a corresponding static solutions  of the  nonlinear
 equation (\ref{s-eq}) with arbitrary values of the parameter $k$.
 We redefine the scalar field as
\begin{equation}\label{trans}
\vp=-\frac 1k\ln \phi\,,\qquad \phi=e^{-k\vp}\,.
\end{equation}
It follows that
 \begin{equation}\label{eq-change}
 \eta^{\mu\nu}\left(\vp_{,\mu\nu}- k\vp_{,\mu}\vp_{,\nu}\right)=
-\frac1{k\phi}\,{ \eta^{\mu\nu}\phi_{,\mu\nu}}=0\,.
 \end{equation}
Thus, under the transformation (\ref{trans}), the nonlinear
equation (\ref{s-eq}) is transformed to the linear one
(\ref{w-eq}). Consequently, the field $\vp$ satisfies the equation
(\ref{s-eq}) if and only if the new field $\phi$ satisfies
(\ref{w-eq}). We use this transformational  property to derive
 certain exact static solutions of (\ref{s-eq}).

Start with a solution of (\ref{w-eq}) of the form
 \begin{equation}\label{test-sol}
 \phi=1-k\,\frac{m}{R}\,, \qquad R=||\bbr-\bbr_0||\,.
 \end{equation}
 Under  (\ref{trans}), it transforms to a  1-singular solution
of (\ref{s-eq})
 \begin{equation}\label{s-sol}
\vp=-\frac 1k \ln{\Big(1-k\,\frac{m}{R}\Big)}\,.
\end{equation}
This solution
 vanishes at infinity as $\vp\to m/R$.
 Also, in the limiting case $k\to 0 $, the solution (\ref{s-sol})
 approaches the
Newtonian potential. Although these limits hold independently of
the sign of $k$, for $k<0$,  (\ref{s-sol}) is singular only at a
point $\bbr=\bbr_0$ and smooth at other points. We will see in
the sequel that precisely the negative values of the parameter $k$
yield the attraction between the singularities.

 Due to the transformation (\ref{trans}),  the solution
 (\ref{w-sol2}) with $N$ singular points also can be transformed to a
 static solution of (\ref{s-eq}). We take it in the following form
\begin{equation}\label{s-sol0}
\vp=-\frac 1k \ln{\Big(1-k\sum^N_{i=1} \frac{m_i}{R_i}\Big)}\,,
\end{equation}
where $\bR_i=\bbr-\bbr_i$ while $\bbr_i$ is a radius vector to
the $i$-th point. Some analytical properties of this solution can
be  motivated from the physical point of view. It is singular at
the points $\bbr=\bbr_i$ and smooth otherwise. In the small
neighborhoods of the singular points, the solution is
approximately spherically symmetric.  This full spherical symmetry is
 reinstated on a long
distance from the set of $N$ singular points, where the solution is
approximated by
\begin{equation}\label{s-sol0x}
\vp=-\frac 1k \ln{\Big(1-k\,\frac{M}{r}\Big)}\,, \qquad
M=\sum^N_{i=1} m_i\,.
\end{equation}
In other words, the masses of the sources are approximately additive.
 Since the solution (\ref{s-sol0}) is static its singularities do
 not interact.
\section{Non-static solution generated}
\subsection{Instantaneous dynamical Lorentz transformations}
 The solutions exhibited above are static. In order to describe an
 interaction between singularities we are going to derive a certain set of
 dynamical solutions with the same type of singularities.
 For this we have to move the singular points on curved trajectories.
 The corresponding field $\vp$ will also move.
For a Lorentz invariant equation the dynamical solution  can be
derived by a corresponding Lorentz transformation.

Let us start with  a unique singular point which moves on a
straight trajectory with a constant velocity $\bv$. This motion
can be represented as a Lorentz transformation of a stationary
point $\bbr$ with the velocity $-\bv$.
 Using the   Lorentz transformation with the velocity
 $-\bv$ we have (see Appendix A)
\begin{equation}\label{inst-lor}
\bbr\to\bbr -\a \bv <\bv,\bbr>-\,\b \bv t\,,
\end{equation}
where the Lorentz parameters are denoted by
\begin{equation}\label{lor-par}
\b=\frac 1 {\sqrt{1-v^2}}\,, \qquad\qquad \a=\frac
1{v^2}\left(1-\frac 1 {\sqrt{1-v^2}}\right)\,.
\end{equation}
To represent a motion of $N$ singular points on distinct  straight
trajectories, we apply, in their neighborhoods,
$N$ independent  Lorentz transformations
\begin{equation}\label{inst-lor-x}
\bbr_i\to\bbr_i -\a_i \bv_i <\bv_i,\bbr_i>-\,\b_i \bv_i t\,,
\end{equation}
with the Lorentz parameters
\begin{equation}\label{lor-par-x}
\b_i=\frac 1 {\sqrt{1-v_i^2}}\,, \qquad\qquad \a_i=\frac
1{v_i^2}\left(1-\frac 1 {\sqrt{1-v_i^2}}\right)\,.
\end{equation}

 In order to construct
a time dependent solution, we apply, for the static solutions, an
instantaneous Lorentz transformations with time dependent
velocities. Every such transformation acts only in a neighborhood
of the corresponding singular point of the mass $m_i$. In
particular, for a curved trajectory $\bpsi(t)$, we substitute in
(\ref{inst-lor-x})
 $\bv$ by $\dbpsi$ and
$\bv t$ by $ \bpsi$. Consequently,  a {\it dynamical
 Lorentz transformation}
which depends on the trajectory is
\begin{equation}\label{inst-lor1}
\bbr \to \bbr -\a\dbpsi<\dbpsi,\bbr> -\b\bpsi\,.
\end{equation}
In order to  deal with a motion of  of $N$ distinct singularities,
 we define a set of instantaneous dynamical Lorentz
 transformations  ($i=1,\cdots,N$)
\begin{equation}\label{inst-lor2}
(\bbr -\bbr_i)\to (\bbr -\bbr_i)-\a_i\dbpsi_i
<\dbpsi_i,(\bbr-\bbr_i)> -\b_i\bpsi_i\,,
\end{equation}
 where the Lorentz parameters $\a_i$ and $\b_i$ are functions
 only of the velocity $||\dbpsi_i||^2$
\begin{equation}\label{lor3}
\b_i=\frac 1 {\sqrt{1-||\dbpsi_i||^2}}\,, \qquad\qquad \a_i=\frac
1{||\dbpsi_i||^2}\left(1-\frac 1 {\sqrt{1-||\dbpsi_i||^2}}\right)\,.
\end{equation}
\subsection{Non-static solution of the linear field equation}
 Now, consider the
 linear field equation
 \begin{equation}\label{eq1}
 \eta^{\mu\nu}\vp_{,\mu\nu}=0\,.
 \end{equation}
 Recall that this equation is a special case, $k=0$, of our non-linear
 model (\ref{s-eq}).

 We start with an exact  static solution of (\ref{eq1})
 \begin{equation}\label{eq2}
 \vp=\sum^N_{i=1}\frac {m_i}{R_i}\,,\qquad \bR_i=\bbr - \bbr_i\,.
 \end{equation}
 The transformation (\ref{inst-lor2})
 leads  to a time depended ansatz
\begin{equation}\label{s-sol3}
\vp=\sum_{i=1}^N \frac {m_i}{R_i}\,,\qquad
\bR_i=(\bbr - \bbr_i)-\a_i\dbpsi_i <\dbpsi_i,(\bbr-\bbr_i)>
-\b_i\bpsi_i\,.
\end{equation}
 Clearly, for an arbitrary function $\bpsi(t)$,  (\ref{s-sol3}) does not
 satisfy (\ref{eq1}).
 Let us  look for  functions $\bpsi_i(t)$ so that
 (\ref{s-sol3}) is a solution of  (\ref{eq1}), at least
 approximately.
 Calculate, approximately, the left hand side of  (\ref{eq1})
 for $\vp$ given in (\ref{s-sol3}).
In particular, we omit the time derivatives of the Lorentz functions
 $\a_i$ and $\b_i$, which  bring only terms
 ${\mathcal O}(\ddbpsi\dbpsi)$.
In this lowest order approximation, cf. Appendix B,
\begin{equation}
\square\, \vp=\sum_{i=1}^N \frac{m_i\b_i}{R_i^3}\,<\ddbpsi_i,\bR_i>
+ {\mathcal O}(\ddbpsi\dbpsi)\,.
\end{equation}
Consequently  $\vp$ is an approximate solution  of the field
equation (\ref{w-eq}) if and only if  $\ddbpsi_i=0$.  Thus, the
linear field equation (\ref{w-eq}) can serve as  a field
 model for a free inertial motion
 of a system of $N$ singularities.

In fact, for $\ddbpsi_i=0$, i.e. for $\bpsi_i=\dbpsi_i t$,
this result is exact.
 Indeed,  at the point $\bR_i=0$
\begin{equation}\label{2-6}
 \bbr - \bbr_i=\dbpsi_i \left(\a_i<\dbpsi_i,(\bbr-\bbr_i)>+\b_i t\right)\,,
\end{equation}
 which results in the equation of motion of a free
particle
\begin{equation}\label{2-9}
\bbr-\bbr_i=\dbpsi_i t\,.
\end{equation}

\subsection{Non-static solution of the non-linear field equation}
Let us now turn  to the full non-linear scalar field
equation
 \begin{equation}\label{eq3}
 \eta^{\mu\nu}\left(\vp_{,\mu\nu}- k\vp_{,\mu}\vp_{,\nu}\right)=0\,.
\end{equation}
If the transformation  (\ref{inst-lor2}) is applied to its
 static solution
\begin{equation}\label{eq4}
\vp=-\frac 1k \ln{\Big(1-k\sum^N_{i=1}
\frac{m_i}{R_i}\Big)}\,,\qquad
 \bR_i=\bbr-\bbr_i\,.
\end{equation}
 we will  obtain no more than the inertial motion $\ddbpsi=0$.
 This is because
 the equation (\ref{eq3}) is Lorentz invariant.
 Instead of that, we start with a corresponding  approximate solution,
\begin{equation}\label{eq4x}
\vp=-\frac 1k \sum^N_{i=1}\ln{\left(1-k
\frac{m_i}{R_i}\right)}\,,\qquad
 \bR_i=\bbr-\bbr_i\,.
\end{equation}
which is  a superposition of $N$ independent exact solutions.

 In order to generate a non-static (interaction) solution, we
apply, in a neighborhood of  every singular point,
 Lorentz transformations with  variable
velocities. Thus we come to
 \begin{equation}\label{eq4xx}
\vp=-\frac 1k \sum^N_{i=1}\ln{\left(1-k
\frac{m_i}{R_i}\right)}\,,\qquad
 \bR_i=(\bbr - \bbr_i)-\a_i\dbpsi_i <\dbpsi_i,(\bbr-\bbr_i)>
-\b_i\bpsi_i\,.
\end{equation}
As in the linear case, for an arbitrary function
$\bpsi(t)$, this ansatz is not a solution
 of (\ref{eq3}). Let us now search for conditions under
 which (\ref{eq4xx}) turns into an approximate solution
of (\ref{eq3}) in the lowest order.

 Calculating for (\ref{eq4xx}) the linear second order derivative
 part of (\ref{eq3}), we get
(see Appendices C and D)  \br\label{2-85} &&\sum^N_{i=1}
\frac {m_i\b_i}{R_i^3}
\cdot\frac{<\bR_i,\ddbpsi_i>}{1-k\frac{m_i}{R_i}}=-k\sum_{i\ne
j}^N\frac {\frac {m_i}{R_i^3}}{1-k\frac{m_i}{R_i}}\, \frac
{\frac
{m_j}{R_j^3}}{1-k\frac{m_j}{R_j}} \Big[<\bR_i,\bR_j>+\nonumber \\
&&<\dbpsi_i,\bR_j>\,<\dbpsi_i,\bR_i>
\,(\b_i\b_j+\a_i+\a_j-\a_i\a_j<\dbpsi_i,\dbpsi_j>)\Big]+
 \mathcal{O}(\dbpsi\ddbpsi)\,.
 \er
 The expressions are calculated
 up to  $\mathcal{O}(\dbpsi\ddbpsi)$.
 To this accuracy,
  the derivatives of the Lorentz parameters
 $\a_i,\b_i$ are zero.

The second line of (\ref{2-85}) is  of the order
 $ \mathcal{O}(\dbpsi_i\dbpsi_j)$.
 For the lowest approximation, we neglect  this term.
 Moreover, to the
 same accuracy, we can take $\b_i=1$.
 Thus, from this stage, our results will be valid only for
 slow motion  of the singularities.

 To sum up, we derived
 that the ansatz (\ref{eq4xx}) is an approximate   solution of the
 field equation (\ref{eq3}) only if the functions $\bpsi_i(t)$
 satisfy
\begin{equation}\label{2-86}
\sum^N_{i=1} \frac {m_i}{R_i^3}
\,\frac{<\bR_i,\ddbpsi_i>}{1-k\frac{m_i}{R_i}}=-k\sum_{i\ne
j}^N\frac {\frac {m_i}{R_i^3}}{1-k\frac{m_i}{R_i}}\, \frac
{\frac {m_j}{R_j^3}}{1-k\frac{m_j}{R_j}} <\bR_i,\bR_j>\,,
\end{equation}
where
 \begin{equation}\label{eq10}
  \bR_i=(\bbr - \bbr_i)-\a_i\dbpsi_i <\dbpsi_i,(\bbr-\bbr_i)>
-\b_i\bpsi_i\,.
\end{equation}

 \section{An interaction is generated}
Both sides of the equation (\ref{2-86}) are functions of a
point $\bbr$. It is singular at $N$ points $\bR_i=0$ and
regular at other points. Let us examine how this relation
can be satisfied in a neighborhood of a singular point.
Consider the $p$-th singularity. We will examine the field
at an arbitrary point $\bbr$ close to this singularity,
i.e., we have $\bR_p\to 0$. For the $i$-th singularity
\begin{equation}\label{2-87}
\bR_p\to 0 \text{\quad yields\quad }\bR_i\to \bR_{ip}\text{\quad
for\quad } i\ne p\,,
\end{equation}
where $\bR_{ip}$ is a vector directed from the point $i$ to the
point $p$.

 On the left hand side of the equation (\ref{2-86}), there is only one
singular term
\begin{equation}\label{si1}
\frac{m_p}{R_p^3}\frac{<\bR_p,\ddbpsi_p>} {1-k\frac{m_p}{R_p}}\,.
\end{equation}
The singular term on the right hand side of (\ref{2-86}) is
\begin{equation}\label{si2}
-k\,\frac {\frac {m_p}{R_p^3}}{1-k\frac{m_p} {R_p}}\sum_{j\ne
p}\frac {\frac {m_j}{R_j^3}} {1-k\frac{m_j}{R_j}}<\bR_p,\bR_j>\,.
\end{equation}
The terms (\ref{si1}) and (\ref{si2}) are of the order ${\mathcal
O}(R^{-2})$ near the $p$-th singularity.

 They cancel
each other if and only if
\begin{equation}\label{2-89}
<\bR_p,\ddbpsi_p>\,=-k\sum_{j\ne p}\frac {\frac
{m_j}{R_j^3}}{1-k\frac{m_j}{R_j}} \,<\bR_p,\bR_j>\,.
\end{equation}
 Take into account that  the point $\bbr$ (in $\bR_p$) is
arbitrary. Hence (\ref{2-89}) is valid only if
\begin{equation}
\ddbpsi_p=-k\sum_{j\ne p}\frac {\frac {m_j}{R_j^3}}
{1-k\frac{m_j}{R_j}}\bR_j\,.
\end{equation}
For the limiting values, we have  in (\ref{2-87})
\begin{equation}\label{2-92x}
\ddbpsi_p=-k\sum_{j\ne p}\frac {m_j}{R_{jp}^3
\left(1-k\frac{m_j}{R_{jp}}\right)}\,\bR_{jp} \,.
\end{equation}
For distances greater than  the Schwarzschild radius $r=km$,
(\ref{2-92x}) remains in the form
 \begin{equation}\label{2-92}
\ddbpsi_p=-k\sum_{j\ne p}\frac {m_j}{R_{jp}^3}\,\bR_{jp}\,.
\end{equation}
For a system of two singular points, it yields
\begin{equation}\label{2-93}
m_1\ddbpsi_1=k\frac {m_1m_2}{R_{12}^3}\,\bR_{12}\,.
\end{equation}
For $k<0$, (\ref{2-92})  and (\ref{2-93}) result in attraction
between the particles. The absolute value of $k$ is unimportant,
since it amounts to the rescaling of the mass.

This way the Newton-type law of attraction is derived from the
scalar field equation.
 \section{Qualitative Description of the Algorithm}
In this section we give  a qualitative description of the proposed
algorithm for deriving the equations of motion from the field
equation. Consider a field equation of a general type
\begin{equation}\label{fe2}
a\Phi_{,\mu\nu}-b\Phi_{,\mu}\Phi_{,\nu}=0
\end{equation}
without specification of the tensorial nature of the field
variable $\Phi$. Here,  the coefficients $a$ and $b$ are the
dimensionless functions of the field variable (or constants)
$$a=a(\Phi),\qquad b=b(\Phi).$$
We do require, however, that (\ref{fe2}) is Lorentz invariant.

Let the field equation (\ref{fe2}) has a static spherical
symmetric solution $\Phi(\bbr-\bbr_0)$ with a singularity
located at $\bbr=\bbr_0$. Denote the Lorentz
transformation based on the velocity $\bv$ by $L(\bv)$.
Consequently, if  $\Phi=\Phi(\bbr-\bbr_0)$ is a time
independent solution of  (\ref{fe2}) then $ L(\bv)\Phi$ is
also a solution of the same equation for an arbitrary
Lorentz transformation $L_(\bv)$. Note that if $\Phi$ is a
multi-component quantity like a tensor then $L(\bv)$
involves not only a coordinate change but, also, a
transformation of components of $\Phi$. The solution $
L(\bv)\Phi$ describes the field of a pointwise singularity
moving with a constant velocity $\bv$ on the trajectory
$\bpsi=\bv t$.

Let us try to construct a generalization of the Lorentz
transformation used above such that the origin moves on a
curved trajectory $\bpsi=\bpsi(t)$. Denote such a
transformation by $N(\bpsi)$. The choice
$$N(\bpsi)=L({\dot\bpsi})$$
is a plausible candidate. Correspondingly, $N(\bpsi)\Phi$
is a rigid motion of the field.

 Now substitute $N(\bpsi)\Phi$ in
(\ref{fe2}). If $\dot \bpsi=const$ then $N(\bpsi)\Phi$ is
also a solution of (\ref{fe2}). If not, then the linear
part produces extra terms. These extra terms come from two
sources:
\begin{itemize}
\item [{\bf(i)}] {\it The derivatives of the Lorentz root}
$\sqrt{1-{|\dot\bpsi|}^2}$. In our consideration the
velocity of the particle $\dot\bpsi$ as well as its time
and spatial derivatives are assumed to be small. It
follows that the derivatives of the root are of the form
${\cal O}(|{\ddot{\bpsi}}|\, |\dot\bpsi|)$. Thus they may
be rejected.
\item [{\bf(ii)}] {\it The linear  part.}
Since $\Phi$ is time independent the first order
derivatives of $\Phi$ are only the spatial ones. Thus, the
first order derivatives  of $L({\dot\bpsi})\Phi(x)$
involve spatial derivatives of $\Phi$ multiplied by
$\dot\bpsi$. The second order derivatives of
$L({\dot\bpsi})\Phi(x)$ cancel each other exactly in the
same way as in a Lorentz transformed solution
$L(\bv)\Phi(x)$. One exception: The second order derivative
$\ddot\psi$ multiplied by spatial first order derivatives
of $\Phi$ do remain. This extra term that comes from the
linear part of the field equation is the {\it agent of
inertia}.
\item [{\bf(iii)}] {\it The quadratic part.}
It involves only first order derivatives of the field.
Consequently the fact that $ \dot\bpsi$ is variable does
not affect it's form.
\end{itemize}
Construct now  a solution that describes a field of $N$
particles, i.e. a field with $N$ singular points. It can
be approximated by a superposition of 1-singular solutions
moving on arbitrary trajectories $\bpsi_j=\bpsi_j(t)$.
$$\Phi=\sum_j L_j(\dot\bpsi_j)\Phi_j(x).$$
Substituting this approximate solution in (\ref{fe2}) we
obtain, in the linear part, only the second derivatives
$\ddot\bpsi_j$ multiplied by the spatial first derivatives
of $\Phi$.

 Consider the quadratic part. It is composed of the
first order derivatives of $L_j(\dot\bpsi_j)\Phi_j(x)$
multiplied by the first order derivatives of
$L_k(\dbpsi_k)\Phi_k(x)$. If $j=k$, since $\Phi_k(x)$ is a
solution of  (\ref{fe2}), these products are cancelled by
the linear part operating   on $\Phi_k(x)$. If $j\ne k$
the products will be declared to be an approximation to
the interaction between the $j$-th and the $k$-th
particles.

 Near the $k$-th singularity, for the linear part,
 only the terms  coming from   $\Phi_k(x)$, similar to (\ref{si1}), will be dominant.
Likewise, for the quadratic part, only the terms involving the
derivatives of $\Phi_k(x)$, similar to (\ref{si2}), will be dominant.
Equating, near the
singularity, of the two terms above (again to the leading order)
should, hopefully, result in the Newton-type law of attraction.
 \section{Conclusions}
The motivation for  our paper is to learn what  properties of
the Einstein field equations lead to derivability the proper equations
of motion for the singularities. We conjecture that 
it pertains to the specific structure of the field equations: 
 As partial differential equations,  they are
linear in the second order derivatives and quadratic in the first
order derivatives of the field variables. Furthermore, we conjecture  that the
quadratic terms are  the {\it  agent of interaction}, they 
have to generate the force expression. The second order derivatives
play a role of the {\it agent of inertia}, it has to generate
the mass-times-acceleration term.

We curry out the above plan for  a
simple non-linear scalar model. It is remarkable that only with
the slow motion assumption, one is ultimately led to the
Newton-type law of attraction. It should be emphasized that our
scheme does not pretend to be more precise  than the  alternative
methods cited above. Instead, we are looking for a proper
identification of the terms of the field equation with the
inertial and interaction terms of the equations of motion.

In order to apply the proposed scheme to Einstein's equation, one
has to solve the problem of  consistency \cite{k-i}. Every one of
ten independent equations $R_{\mu\nu}=0$ has  to give the same
equation of motion. Also the Bianchy identities have to be taken
into account and their role in our scheme has to be revealed. It
is well known that the Bianchy identities play a crucial role in
EIH-procedure.

An other extension is in the direction of embedding \cite{k-i}.
The trajectories obtained are approximate. At this point, two
avenues are open. The first one, which is adopted by EIH is to
get higher order approximations to the trajectories. This
procedure is also used in the PPN approach. By these methods, the
successive approximations become highly singular near the particle
trajectories. The second avenue is to embed the singularities in
a field satisfying the field equations. For that purpose, the
successive approximations should add regular terms (and,
possible, low order singular terms) near the trajectories.

The method proposed in this paper can hopefully be useful  for a
rigorous proof of the geodesic postulate.
\section*{Acknowledgment}
We are grateful to the anonymous  referee for most careful
reading of our manuscript. His valuable  comments and remarks as
well as his guide to the relevant literature  were very helpful.

\begin{appendix}
\section{Lorentz transformations}
Lorentz transformations are usually written
 in a very special form when the axes of two reference systems
 are parallel one to the other. In particular, when a reference  system
 $\{\tilde{t},\tilde{x},\tilde{y},\tilde{z}\}$
 moves relative to another reference system $\{t,x,y,z\}$
 with a velocity parallel to the
 $x$-axis, the corresponding Lorentz transformation is
\begin{equation} \label{lor-1}
\tilde{x}=\b(x+vt), \quad \tilde{y}=y,
 \quad \tilde{z}=z, \quad
 {\tilde{t}}=\b (t+vx) \,,
\end{equation}
with the Lorentz parameter
\begin{equation} \label{lor-2}
\b=\frac 1{\sqrt{1-v^2}}\,.
\end{equation}
 Recall that we use a system of units with $c=1$.

 Since the different directed Lorentz transformations do not commute,
 a general transformation (with an arbitrary directed vector of velocity)
 can not be generated by a successive application of three orthogonal
 transformations (relative to three axes).  Although, a formula for the
 Lorentz transformation with a general velocity vector is known from the
 literature, we present here a  brief derivation from
 (\ref{lor-1}). This derivation is instructive  because  the proof
 conforms with the method used in this article.

Consider a reference system which axes are parallel
 to the corresponding
axes of a rest reference system. Let the origin of the   reference
system move with an arbitrary directed velocity $\mathbf{v}$.
Consider a radius vector $\bbr$ directed
 to an arbitrary point in space. Its projection
 on the direction of $\bv$ is
\begin{equation} \label{proj}
P_{\bv}\bbr=\bv\,\frac{<\bv,\bbr>}{v^2}\,.
\end{equation}
Exhibit the vector $\bbr$ as a sum of its tangential and
 normal parts
\begin{equation} \label{}
\bbr=P_{\bv}\bbr+N_{\bv}\bbr\,, \qquad
 N_{\bv}\bbr=\bbr-\bv\,\frac{<\bv,\bbr>}{v^2}\,.
\end{equation}
Due to (\ref{lor-1}), for a Lorentz transformation
 directed by $\bv$, these two parts transform  as
\br
P_{\bv}{\tilde {\mathbf{r}}}&=&\b(P_{\bv}\bbr +\bv t)\,,\\
N_{\bv}{\tilde {\mathbf{r}}}&=&N_{\bv}\bbr\,. \er
Consequently, the transform of the spatial coordinates is
 \begin{equation} \label{}
{\tilde {\mathbf{r}}}=P_{\bv}{\tilde {\mathbf{r}}}+
 N_{\bv}{\tilde {\mathbf{r}}}=
 \b (P_{\bv}\bbr +\bv t)+N_{\bv}\bbr\,,
\end{equation}
or, explicitly,
 \begin{equation} \label{lor1n}
{\tilde {\mathbf{r}}}=\bbr+
 \bv\left(\b t-\a<\bbr,\bv>\right)\,,
\end{equation}
 where
 \begin{equation} \label{lor1x}
 \a=\frac{(1-\b)}{v^2}=\frac 1{v^2}\left(1-
 \frac 1 {\sqrt{1-v^2}}\right)\,.
 \end{equation}
The change of the time coordinate is also governed only
 by the
tangential part of the vector $\bbr$
 \begin{equation} \label{}
\tilde{t}=\b(t+||P_{\bv}\bbr||v)=
 \b\left(t+<P_{\bv}\bbr,\bv>\right)\,,
\end{equation}
or, explicitly,
\begin{equation} \label{lor2n}
\tilde{t}=\b\left(t+<\bbr,\bv>\right).
\end{equation}
In the special case of a motion parallel to the axis $x$, the
relations (\ref{lor1n}), (\ref{lor2n})  reduce to the ordinary
form of the Lorentz transformation (\ref{lor-1}). Therefore an
arbitrary  Lorentz transformation takes the form
\begin{equation} \label{a12}
\left\{\begin{array}{ll}
&\tilde{t}=\b\left(t+<\bbr,\bv>\right)\,,\\
&{\tilde {\mathbf{r}}}=
 \bbr+\bv\left(\b t-\a<\bbr,\bv>\right)\,.
\end{array}\right.
\end{equation}

\section{The linear equation: The 1-point solution moved}
Calculate the d'Alembertian of the field $\vp$ of a single
 particle moving on a trajectory $\bpsi(t)$
\begin{equation}\label{B1}
\vp=\frac m{R},\qquad R=||\bR||\,,
\end{equation}
where
\begin{equation}\label{B2}
\bR=(\bbr - \bbr_0)-\a\dbpsi <\dbpsi,
 (\bbr-\bbr_0)>-\b \bpsi\,.
\end{equation}
Here $\a$ and $\b$ are Lorentz functions
 of the velocity $||\dbpsi||^2$
\begin{equation}\label{B3}
\b=\frac 1 {\sqrt{1-||\dbpsi||^2}}\approx 1+\frac 12
||\dbpsi||^2+\frac 38 ||\dbpsi||^4+\cdots
\end{equation}
and
\begin{equation}\label{B3x}
 \a=\frac 1{||\dbpsi||^2}
 \left(1-\frac 1 {\sqrt{1-||\dbpsi||^2}}\right)\approx -
 \frac 12\left(1-\frac 32 ||\dbpsi||^2-\frac 58 ||\dbpsi||^4+\cdots\right)\,.
\end{equation}
The time derivatives of these functions  are correspondingly 
\begin{equation}\label{B3xx}
\frac {d\b}{dt}=2\b'<\dbpsi,\ddbpsi>=<\dbpsi,\ddbpsi>\left(1+\frac
32||\dbpsi||^2+\cdots\right)\,,
\end{equation}
and
\begin{equation}\label{B3xxx}
\frac{d\a}{dt}=2\a'<\dbpsi,\ddbpsi>=-<\dbpsi,\ddbpsi>\left(\frac
32+\frac 54||\dbpsi||^2 + \cdots\right) \,,
\end{equation}
 where $\a', \b'$ denote the derivatives relative to the variable
$\dbpsi$. 
 
 The first order time derivative of
the vector (\ref{B2}) is
\br {\dot \bR}&=&-\left(2\a'\dbpsi<\dbpsi,\ddbpsi>+
 \a\ddbpsi\right)<\dbpsi,\bbr-\bbr_0>\nonumber\\
&&-\a\dbpsi <\ddbpsi,\bbr-\bbr_0> -2\b'\bpsi<\dbpsi,\ddbpsi>  -\b
\dbpsi\,. \er
 Due to (\ref{B3}-\ref{B3xxx}), up to the order
${\mathcal O}(\dbpsi\ddbpsi)$, the second order time
derivative of the field  is \br \frac
{\partial^2\vp}{\partial t^2}
&=&\frac{m\b}{R^3}\left(<\ddbpsi ,\bR>+<\dbpsi
,\bR_t>\right) -3\frac{m\b}{R^5}<\dbpsi
,\bR><\dot{\bR},\bR>\,. \er Thus we get \br \frac
{\partial^2\vp}{\partial t^2} &=&\frac{m\b}{R^3}\left(
<\ddbpsi ,\bR>-\b \,||\dbpsi||^2\right)
 \ + \ 3\frac{m\b^2}{R^5}<\dbpsi ,\bR>^2\nonumber\\
&=&\frac{m\b}{R^3}<\ddbpsi ,\bR> \
+ \  \ \frac{m\b^2}{R^5}\left(3<\dbpsi ,\bR>^2-R^2||\dbpsi||^2\right).
\er

Let us calculate now  the spatial derivatives
of the field (\ref{B1}). Introduce a set of unit
 vectors $\e_1, \e_2,\e_3$ directed along the axes $x,y,z$
 correspondingly. Thus the $x$-component of the vector
 (\ref{B2}) can be written as
 \begin{equation}
\bR_x=\e_1-\a \dbpsi<\dbpsi,\e_1>\,.
\end{equation}
 Consequently,
 the second order derivative is
\begin{equation}
\frac {\partial^2\vp}{\partial
x^2}=-\frac{m}{R^3}||\bR_x||^2 \ + \
 3\frac{m}{R^5}<\bR_x,\bR>^2\,.
\end{equation}
 Due to the spherical symmetry of the field $\vp$,
 its Laplacian takes the form
\begin{equation}
\triangle \vp= 3\frac{m}{R^5}\left(<\bR_x,\bR>^2+
 <\bR_y,\bR>^2+<\bR_z,\bR>^2\right)-
\frac{m}{R^3}\left(||\bR_x||^2+||\bR_y||^2+
 ||\bR_z||^2\right)\,.
\end{equation}
Since
\begin{equation}
<\bR_x,\bR>\,=\,<\bR,\e_1>-\a<\dbpsi,\e_1><\dbpsi,\bR>
\end{equation}
we get
\begin{equation}
<\bR_x,\bR>^2+<\bR_y,\bR>^2+<\bR_z,\bR>^2=
 R^2-2\a<\dbpsi,\bR>^2+
 \a^2||\dbpsi||^2<\dbpsi,\bR>^2\,,
\end{equation}
and
\begin{equation}
||\bR_x||^2+||\bR_y||^2+
 ||\bR_z||^2=3-2\a ||\dbpsi||^2+
 \a^2||\dbpsi||^4\,.
\end{equation}
Thus the Laplacian of the field $\vp$ takes
 the form
\begin{equation}
\triangle \vp=\frac{m}{R^5}(\a^2||\dbpsi||^2-2\a)
 (3<\dbpsi,\bR>^2-||\dbpsi||^2R^2)\,.
\end{equation}
Correspondingly, the d'Alembertian of the field is
\begin{equation}
\square \,\vp=\frac{m\b}{R^3}<\ddbpsi ,\bR>+
 \frac{m}{R^5}(\b^2+2\a-\a^2||\dbpsi||^2)
 \Big(3<\dbpsi ,\bR>^2-R^2||\dbpsi||^2\Big)\,.
\end{equation}
Using the expressions (\ref{B3}) for the
 functions $\a,\b$, we derive
 \begin{equation}
\b^2+2\a-\a^2||\dbpsi||^2=0\,.
\end{equation}
 Consequently, the d'Alembertian of the 1-singular
 ansatz (\ref{B1}) takes the form
\begin{equation}
\square \,\vp=\frac{m\b}{R^3}<\ddbpsi ,\bR>\,.
\end{equation}
 It is clear that a singularity that moves on a straight trajectory with a
 constant velocity $\ddbpsi=0$ is described by a solution of the linear
 equation $\square \,\vp=0.$
\section {The non-linear equation: The motion of a 1-point singularity}
In this section we substitute the ansatz with one singular point
---
\begin{equation}
\vp=-\frac 1k \ln{\left(1-k \frac{m}{R}\right)}\,,\qquad
 \bR=(\bbr - \bbr)-\a\dbpsi<\dbpsi,(\bbr-\bbr)>
-\b\bpsi\,.
\end{equation}
into the left hand side of the non-linear field equation
(\ref{eq3}).  From the calculations above, the time derivatives
of the vector $\bR$ are
\begin{equation}
\bR_t=-\b\dbpsi\,, \qquad \bR_{tt}=-\b\ddbpsi\,.
\end{equation}
Thus  the second order time derivative is \br
\vp_{tt}=\frac{m\b}{R^5}\,\frac{3\b<\bR,\dbpsi>^2+R^2<\bR,\ddbpsi>-\b
R^2||\dbpsi||^2} {1-k\frac{m}{R}}+ \frac{m^2\b^2
k}{R^6}\,\frac{<\bR,\dbpsi>^2}{(1-k\frac{m}{R})^2}. \er As
for the spatial derivatives we use
\begin{equation}
\bR_x=\e_1-\a\dbpsi<\dbpsi,\e_1>\,, \qquad \bR_{xx}=0\,,
\end{equation}
where $\e_1$ is a unit vector along the $x$ axis.
Consequently,
 \br \triangle \vp&=&3\frac
m{R^5}\,\frac{<\bR,\bR_x>^2+<\bR,\bR_y>^2+<\bR,\bR_z>^2}
{1-k\frac{m}{R}}-\nonumber\\
&&\frac
m{R^3}\,\frac{<\bR_x,\bR_x>+<\bR_y,\bR_y>+<\bR_z,\bR_z>}
{1-k\frac{m}{R}}+\nonumber\\
&&\frac
{km^2}{R^6}\,\frac{<\bR,\bR_x>^2+<\bR,\bR_y>^2+<\bR,\bR_z>^2}
{(1-k\frac{m}{R})^2}\,. \er Substituting here the value of
$\bR_x$ we get \br \triangle \vp&=&3\frac m{R^5}\,\frac
{R^2-2\a<\dbpsi,\bR>^2+\a^2||\dbpsi||^2<\dbpsi,\bR>^2}{1-k\frac{m}{R}}-
\frac m{R^3}\,\frac{3-2\a||\dbpsi||^2
+\a^2||\dbpsi||^4}{1-k\frac{m}{R}}+\nonumber\\&&\frac
{km^2}{R^6}\,\frac{R^2-2\a<\dbpsi,\bR>^2+\a^2||\dbpsi||^2<\dbpsi,\bR>^2}
{(1-k\frac{m}{R})^2}\,. \er Thus
 \br \square\,
\vp&=&\frac{m\b}{R^5}\,\frac{3\b<\bR,\dbpsi>^2+R^2<\bR,\ddbpsi>-\b
R^2||\dbpsi||^2}{1-k\frac{m}{R}}+ \frac{m^2\b^2
k}{R^6}\,\frac{<\bR,\dbpsi>^2}{(1-k\frac{m}{R})^2}-\nonumber\\&&
3\frac m{R^5}\,\frac
{R^2-2\a<\dbpsi,\bR>^2+\a^2||\dbpsi||^2<\dbpsi,\bR>^2}{1-k\frac{m}{R}}+
\frac m{R^3}\,\frac{3-2\a||\dbpsi||^2
+\a^2||\dbpsi||^4}{1-k\frac{m}{R}}- \nonumber\\&&\frac
{km^2}{R^6}\,\frac{R^2-2\a<\dbpsi,\bR>^2+
\a^2||\dbpsi||^2<\dbpsi,\bR>^2}{(1-k\frac{m}{R})^2}\nonumber\\
&=&\frac {m\b}{R^3}\,\frac{<\bR,\ddbpsi>}{1-k\frac{m}{R}}+
\frac
m{R^5}(\b^2+2\a-\a^2||\dbpsi||^2)\,\frac{3<\bR,\dbpsi>^2-
R^2||\dbpsi||^2}{1-k\frac{m}{R}}+\nonumber\\&& \frac
{m^2k}{R^6}\,\frac{-R^2+(\b^2+2\a-\a^2||\dbpsi||^2)<\bR,\dbpsi>^2}
{(1-k\frac{m}{R})^2}\,. \er Using the relation
\begin{equation}\label{rel}
\b^2=\a^2||\dbpsi||^2-2\a
\end{equation}
 we obtain
\begin{equation}\label{lead}
\square \, \vp=\frac
{m\b}{R^3}\,\frac{<\bR,\ddbpsi>}{1-k\frac{m}{R}}-\frac
{m^2k}{R^4}\frac 1 {(1-k\frac{m}{R})^2}\,.
\end{equation}
As for the quadratic part of the field equation, to the
same accuracy,
 \br
\eta^{ab}\vp_{,a}\vp_{,b}&=&\frac
{m^2\b^2}{R^6}\,\frac{<\bR,\dbpsi>^2}{(1-k\frac{m}{R})^2}-
\frac {m^2}{R^6}\,\frac{<\bR,\bR_x>^2+<\bR,\bR_y>^2+
<\bR,\bR_z>^2}{(1-k\frac{m}{R})^2}\nonumber\\
&=&\frac
{m^2\b^2}{R^6}\,\frac{<\bR,\dbpsi>^2}{(1-k\frac{m}{R})^2}-
\frac
{m^2}{R^6}\,\frac{R^2-2\a<\dbpsi,\bR>^2+\a^2||\dbpsi||^2
<\dbpsi,\bR>^2}{(1-k\frac{m}{R})^2}\nonumber\\
&=&-\frac {m^2}{R^4}\frac 1 {(1-k\frac{m}{R})^2}+ \frac
{m^2}{R^6}(\b^2+2\a-\a^2||\dbpsi||^2)
\frac{<\bR,\dbpsi>^2}{(1-k\frac{m}{R})^2}\,. \er
 Using once more the
relation (\ref{rel}) we get
\begin{equation}\label{quad}
\eta^{ab}\vp_{,a}\vp_{,b}=-\frac {m^2}{R^4}\frac 1
{(1-k\frac{m}{R})^2}\,.
\end{equation}
When (\ref{lead}) and (\ref{quad}) are  substituted into the left
hand side of  the field equation (\ref{eq3}) we obtain
\begin{equation}\label{lead+quad}
\square \, \vp-k\eta^{ab}\vp_{,a}\vp_{,b}=\frac
{m\b}{R^3}\,\frac{<\bR,\ddbpsi>}{1-k\frac{m}{R}}\,.
\end{equation}
Consequently,  the field  equation (\ref{eq3}) is satisfied  only for a
singularity that moves on a straight trajectory with a constant
velocity.
\section {The non-linear equation. The $N$-point solution moved }
Take the field of  $N$ singular points as a superposition
 \begin{equation}\label{eq4xxap}
\vp=-\frac 1k \sum^N_{i=1}\ln{\left(1-k
\frac{m_i}{R_i}\right)}\,,\qquad
 \bR_i=(\bbr - \bbr_i)-\a_i\dbpsi_i <\dbpsi_i,(\bbr-\bbr_i)>
-\b_i\bpsi_i\,.
\end{equation}
Substitute it in the left hand side of the non-linear field equation
(\ref{eq3}). Since this ansatz a superposition of $N$ independent
solutions, the linear part of the field equation is a sum of the
expressions given in (\ref{lead})
\begin{equation}\label{lead-4}
\square \, \vp=\sum^N_{i=1}\frac
{m_i\b_i}{R_i^3}\,\frac{<\bR_i,\ddbpsi_i>}{1-k\frac{m_i}{R_i}}-\frac
{m_i^2k}{R_i^4}\frac 1 {(1-k\frac{m_i}{R_i})^2}\,.
\end{equation}
  Calculate now the non-linear part
\begin{equation}
\vp_t=-\sum^N_{i=1}\Big(\frac
{<\bR_{it},\bR_i>}{1-k\frac{m_i}{R_i}}\, \frac
{m_i}{R_i^3}\Big)= \sum^N_{i=1}\Big(\frac
{<\dbpsi_i,\bR_i>}{1-k\frac{m_i}{R_i}}\, \frac
{m_i\b_i}{R_i^3}\Big)\,.
\end{equation}
Thus
\begin{equation}
(\vp_t)^2=\sum^N_{i,j=1}\frac {\frac
{m_i\b_i}{R_i^3}}{1-k\frac{m_i}{R_i}}\,\frac {\frac
{m_j\b_j}{R_j^3}}{1-k\frac{m_j}{R_j}}<\dbpsi_i,\bR_i>
<\dbpsi_j,\bR_j>\,.
\end{equation}
As for the spatial derivatives
 \br \label{2-79}
<\nabla \vp,\nabla \vp>&=&\sum^N_{i,j=1}\frac {\frac
{m_i}{R_i^3}} {1-k\frac{m_i}{R_i}}\,\frac {\frac
{m_j}{R_j^3}}{1-k\frac{m_j}{R_j}}
\Big(<\bR_{ix},\bR_i><\bR_{jx},\bR_j>+\nonumber\\
&&<\bR_{iy},\bR_i><\bR_{jy},\bR_j>+
<\bR_{iz},\bR_i><\bR_{jz},\bR_j>\Big)\,. \er
 Applying the relation
\begin{equation}
\bR_{ix}=\e_1-\a_i\dbpsi_i<\dbpsi_i,\e_1>\,,
\end{equation}
 the expression
in the  brackets of (\ref{2-79}) takes the form \br
\Big(\cdots\Big)&=&<\bR_i,\bR_j>-\a_j\Big(<\dbpsi_j,\bR_i>
<\dbpsi_j,\bR_j>\Big)-
\a_i\Big(<\dbpsi_i,\bR_j><\dbpsi_i,\bR_i>\Big)+\nonumber\\
&&\a_i\a_j\Big(<\dbpsi_i,\bR_i> <\dbpsi_i,\dbpsi_j>
<\dbpsi_j,\bR_j>\Big). \er Consequently, the quadratic
part  of the field equation takes the form
 \br\label{quad-fin}
k\eta^{ab}\vp_{,a}\vp_{,b}&=&-k\sum^N_{i,j=1}\frac {\frac
{m_i}{R_i^3}}{1-k\frac{m_i}{R_i}}\quad \frac {\frac
{m_j}{R_j^3}}{1-k\frac{m_j}{R_j}}
\Big[<\bR_i,\bR_j>+\nonumber\\
&&<\dbpsi_i,\bR_j><\dbpsi_i,\bR_i> \Big(\b_i\b_j+
\a_i+\a_j-\a_i\a_j<\dbpsi_i,\dbpsi_j>\Big)\Big] \er
\end{appendix}
Observe that, for $i=j$, this expression coincides with
(\ref{quad}). For $i\ne j$, the expressions in the
 second line of (\ref{quad-fin})
 are not
canceled, they can, however, be neglected in the lowest approximation.



\begin{thebibliography}{999}
\bibitem{Birkbook} F.W.~Hehl and Yu.N.~Obukhov, {\it
    Foundations of Classical Electrodynamics: Charge, Flux, and
    Metric} Birkh\"auser: Boston, MA, (2003).
\bibitem{Itin:2004qr}
  Y.~Itin and F.~W.~Hehl,
  Annals Phys.\  {\bf 312}, 60 (2004).



\bibitem{EIH0} A.~Einstein and J. Grommer, Sitzer. deut. Akad.
Wiss. Berlin, {\bf{2}}  (1927).


\bibitem{EIH1} A.~Einstein, L.~Infeld and B.~Hoffmann,
Annals Math. {\bf{39}}, 65  (1938).

 \bibitem{EIH2} A.~Einstein and L.~Infeld,
 Annals Math. {\bf{41}}, 455 (1940). 

 \bibitem{inf-wal} L. Infeld and P.R. Wallace, Phys. Rev. {\bf{57}} 797, (1940)

\bibitem{EIH3} A.~Einstein and L.~Infeld,
Canad. J. Math., {\bf{1}}, 209 (1949). 

\bibitem{infeld1} L.~Infeld and A.Schild, Revs. Mod. Phys.,
{\bf{21}},  408 (1949). 

\bibitem{infeld2} L.~Infeld, Revs. Mod. Phys., {\bf{29}},
398 (1957).

 \bibitem{Weyl} H. Weyl, {\it Raum, Zeit, Materie} Springer-Verlag, Berlin, 1921, 4th ed., Sec. 36;
 in more detail in the 5th ed. (1923).

\bibitem{edd} A.S. Eddington, {\it The mathematical theory of relativity},
 Cambridge University Press (New York) (1923).

 \bibitem{Lan} C.~Lanczos, Zeitschrift f. Physik, {\bf{44}} 773 (1927).

\bibitem{mat}  M.~Mathisson, Zeitschrift f. Physik {\bf 67} (1931) 270; ibid. {\bf{67}} (1931) 826.

 \bibitem{h-g} P.~Havas and J.N.~Goldberg, Phys.\ Rev. {\bf 128}, 398 (1962).
 
 \bibitem{havas} P.~Havas, Proc. Int. School of Phys. E. Fermi LXVII (1979) p. 74,
 Ed. J. Ehlers, North-Holland.

\bibitem{Fo} B. Fock,
 JETP, {\bf{9}}, 375 (1939).

\bibitem{Petrova} N. Petrova, J. Phys. (USSR) {\bf{19}}, 989 (1949)



\bibitem{Pap} A. Papapetrou, Proc. Phys. Soc (London) {\bf{209}},
248 (1951).

 \bibitem{Car} M. Carmeli, Phys. Lett. {\bf 9}, 132 (1964).


 \bibitem{Dam} T.~Damour and N.~Deruelle,
 Ann. Inst. H. Poincare,
(Phys. Theor.)\textbf{44}, 263 (1986).

\bibitem{Damour:1986ny}
  T.~Damour,
In Proc. of Conf. 300 Years of Gravity, Cambridge, England
(1987).


\bibitem{Blanchet:1998vx}
  L.~Blanchet, G.~Faye and B.~Ponsot,
  Phys.\ Rev.\ D {\bf 58}, 124002 (1998).


\bibitem{And}
  J.~L.~Anderson,
  Phys.\ Rev.\ D {\bf 36}, 2301 (1987).

 \bibitem{Schiff} L.~I. Schiff,
Proc. Natl. Acad. Sci. USA, \textbf{46}, 871 (1960).

 \bibitem{Anderson:2005vj}
  J.~L.~Anderson,
  arXiv:gr-qc/0511093.


\bibitem{E-G}
  J.~Ehlers and R.~Geroch,
  Annals Phys.\  {\bf 309}, 232 (2004).

\bibitem{St} S. Sternberg,
Proc. Natl. Acad. Sci. USA, \textbf{96}, 8845 (1999).

 
\bibitem{F-B} Y.~Foures-Bruhat
 Acta. Math. \textbf{88}, 141 (1952).

 \bibitem{WM} K. Watt, C. W. Misner,
 gr-qc/9910032.
 \bibitem{k-i}
S. Kaniel, Y. Itin,
  arXiv:gr-qc/0101011.


\end{thebibliography}
\end{document}